\documentclass[times,twocolumn,final,authoryear]{elsarticle}

\usepackage{prletters}
\usepackage{framed,multirow}
\usepackage{balance}

\usepackage{amsmath}
\usepackage{amssymb}
\usepackage{latexsym}
\usepackage{mathtools}
\usepackage{nicefrac}

\usepackage[british,english]{babel}

\usepackage{array}
\usepackage{booktabs}

\usepackage[algoruled,vlined,linesnumbered]{algorithm2e}

\usepackage{url}
\usepackage{xcolor}
\definecolor{newcolor}{rgb}{.8,.349,.1}

\usepackage{tikz}
\usepackage{pgfplots}
\pgfplotsset{compat=1.3}
\usepackage{filecontents}
\usetikzlibrary{matrix}
\usetikzlibrary{calc}
\usetikzlibrary{arrows}
\usetikzlibrary{patterns}
\usetikzlibrary{plotmarks}
\usetikzlibrary{intersections} 
\usetikzlibrary{decorations.pathmorphing}
\usetikzlibrary{decorations.pathreplacing}
\usetikzlibrary{decorations.markings,arrows} 
\usetikzlibrary{decorations.shapes}
\usetikzlibrary{fit}
\usetikzlibrary{backgrounds}
\usetikzlibrary{trees}

\newcommand{\set}[1]{\left\{ #1 \right\}}

\newcommand{\dataset}{\mathcal{D}}
\newcommand{\loss}[1]{\mathcal{L}\left( #1 \right)}
\newcommand{\closs}[1]{\mathcal{L}_{\text{c}}\left( #1 \right)}
\newcommand{\mloss}[1]{\mathcal{L}_{\text{m}}\left( #1 \right)}
\newcommand{\texta}{C^{\alpha}}
\newcommand{\sorteda}{C^{\alpha'}}
\newcommand{\textb}{C^{\beta}}
\newcommand{\sortedb}{C^{\beta'}}

\DeclareMathOperator*{\median}{median}
\DeclareMathOperator*{\mean}{mean}
\DeclarePairedDelimiter{\ceil}{\lceil}{\rceil}
\DeclarePairedDelimiter{\floor}{\lfloor}{\rfloor}

\journal{Pattern Recognition Letters}

\begin{document}

\ifpreprint
  \setcounter{page}{1}
\else
  \setcounter{page}{1}
\fi

\begin{frontmatter}

\title{Representation learning for very short texts using weighted word embedding aggregation}

\author[1]{Cedric \snm{De Boom}
		  } 
\ead{cedric.deboom@ugent.be}
\author[1]{Steven \snm{Van Canneyt}}
\author[1]{Thomas \snm{Demeester}}
\author[1]{Bart \snm{Dhoedt}}

\address[1]{Ghent University -- iMinds, Department of Information Technology, Technologiepark 15, 9052 Zwijnaarde, Belgium}

\received{1 May 2013}
\finalform{10 May 2013}
\accepted{13 May 2013}
\availableonline{15 May 2013}
\communicated{Cedric De Boom}

\begin{abstract}
Short text messages such as tweets are very noisy and sparse in their use of vocabulary.
Traditional textual representations, such as tf-idf, have difficulty grasping the semantic meaning of such texts, which is important in applications such as event detection, opinion mining, news recommendation, etc.
We constructed a method based on semantic word embeddings and frequency information to arrive at low-dimensional representations for short texts designed to capture semantic similarity.
For this purpose we designed a weight-based model and a learning procedure based on a novel median-based loss function.
This paper discusses the details of our model and the optimization methods, together with the experimental results on both Wikipedia and Twitter data.
We find that our method outperforms the baseline approaches in the experiments, and that it generalizes well on different word embeddings without retraining.
Our method is therefore capable of retaining most of the semantic information in the text, and is applicable out-of-the-box.
\end{abstract}


\end{frontmatter}


\section{Introduction}
\label{sec:introduction}
Short pieces of texts reach us every day through the use of social media such as Twitter, newspaper headlines, and texting.
Especially on social media, millions of such short texts are sent every day, and it quickly becomes a daunting task to find similar messages among them, which is at the core of applications such as event detection (\cite{DeBoom:2015ur}), news recommendation (\cite{Jonnalagedda:2013ie}), etc.

In this paper we address the issue of finding an effective vector representation for a very short text fragment.
By effective we mean that the representation should grasp most of the semantic information in that fragment.
For this we use semantic word embeddings to represent individual words, and we learn how to weigh every word in the text through the use of tf-idf (term frequency - inverse document frequency) information to arrive at an overall representation of the fragment.

These representations will be evaluated through a semantic similarity task.
It is therefore important to point out that textual similarity can be achieved on different levels.
At the most strict level, the similarity measure between two texts is often defined as being (near) paraphrases.
In a more relaxed setting one is interested in topic- and subject-related texts.
For example, if a sentence is about the release of a new Star Wars episode and another about Darth Vader, they will be dissimilar in the most strict sense, although they share the same underlying subject.
In this paper we focus on the broader concept of topic-based semantic similarity, as this is often applicable in the already mentioned use cases of event detection and recommendation.

Our main contributions are threefold.
First, we construct a technique to calculate effective text representations by weighing word embeddings, for both fixed- and variable-length texts.
Second, we devise a novel median-based loss function to be used in the context of minibatch learning to mitigate the negative effect of outliers.
Finally we create a dataset of semantically related and non-related pairs of text from both Wikipedia and Twitter, on which the proposed techniques are evaluated.

We will show that our technique outperforms most of the baselines in a semantic similarity task.
We will also demonstrate that our technique is independent of the word embeddings being used, so that the technique is directly applicable and thus does not require additional model training when used in different contexts, in contrast to most state-of-the art techniques.

In the next section, we start with a summary of the related work, and our own methodology will be devised in Section \ref{sec:methodology}.
Next we explain how data is collected in Section \ref{sec:datacollection}, after which we discuss our experimental results in Section \ref{sec:experiments}.

\section{Related work}
\label{sec:relatedwork}
In this work we use so-called word embeddings as a basic building block to construct text representations.
Such an embedding is a distributed vector representation of a single word in a fixed-dimensional semantic space, as opposed to term tf-idf vectors, in which a word is represented by a one-hot vector (\cite{Achananuparp:2008uf, Manning:2009uf}).
A word's term frequency (tf) is the number of times the word occurs in the considered document, and a word's document frequency (df) is the number of documents in the considered corpus that contain that word.
Its (smoothed) inverse document frequency (idf) is defined as:
\begin{align}
\text{idf} \triangleq \log\frac{N}{1+\text{df}},
\end{align}
in which $N$ is the number of documents in the corpus (\cite{Manning:2009uf}).
A tf-idf-based similarity measure is based on exact word overlap.
As texts become smaller in length, however, the probability of having words in common decreases.
Furthermore, these measures ignore synonyms and any semantic relatedness between different words, and are prone to negative effects of homonyms.

Instead of relying on exact word overlap, one can incorporate semantic information into the similarity process.
Latent Semantic Indexing (LSI) and Latent Dirichlet Allocation (LDA) are two examples, in which every word is projected into a semantic (topic) space (\cite{Deerwester:1990gu, Blei:2003tn}).
At test time, inference is performed to obtain a semantic vector for a particular sentence.
Both training and inference of standard LSI and LDA, however, are computationally expensive on large vocabularies.

Although LSI and LDA have been used with success in the past, Skip-gram models have been shown to outperform them in various tasks (\cite{Mikolov:2013wc, Lebret:2015wo}).
In Skip-gram, part of Google's word2vec toolkit\footnote{\label{footnote:word2vec}Available at \url{code.google.com/archive/p/word2vec}}, distributed word embeddings are learned through a neural network architecture to predict its surrounding words in a fixed window.

Once the word embeddings are obtained, we have to combine them into a useful sentence representation.
One possibility is to use an multilayer perceptron (MLP) with the whole sentence as an input, or a 1D convolutional neural network
(\cite{Collobert:2011tk, Hu:2014uo, Xu:2015vc, Johnson:2015vc}).
Such an approach, however, requires either an input of fixed length or aggregation operations -- such as dynamic k-max pooling (\cite{Kalchbrenner:2014wl}) -- to arrive at a sentence representation that has the same dimensionality for every input.
Recurrent neural networks (RNNs) and variants
can overcome the problem of fixed dimensionality or aggregation, since one can feed word after word in the system and in the end arrive at a text representation (\cite{Sutskever:2014ty, Sordoni:2015uj, Sundermeyer:gj}).
The recently introduced Skip-thought vectors, heavily inspired on Skip-gram, combine the learning of word embeddings with the learning of a useful sentence representation using an RNN encoder and decoder (\cite{Kiros:2015uq}).
RNN-based methods present a lot of advantages over MLPs and convolutional networks, but still retraining is required when using different types of embeddings.


Paragraph2vec is another method, inspired by the Skip-gram algorithm, to derive sentence vectors (\cite{Le:2014vd}).
The technique requires the user to train vectors for frequently occurring word groups.
The method, however, is not usable in a streaming or on-the-fly fashion, since it requires retraining for unseen word groups at test time.

Aggregating word embeddings through a mean, max, min\dots~function is still one of the most easy and widely used techniques to derive sentence embeddings, often in combination with an MLP or convolutional network (\cite{Weston:2014tb, dosSantos:2014tr, Yin:2015vz, Collobert:2011tk}).
On one hand, the word order is lost, which can be important in e.g.~paraphrase identification.
On the other hand, the methods are simple, out-of-the-box and do not require a fixed length input.

Related to the concepts of semantic similarity and weighted embedding aggregation, there is extensive literature.
\cite{Kusner:2015vc}~calculate a similarity metric between documents based on the travel distance of word embeddings from one document to another one.
We on the other hand will derive vectors for the documents themselves.
\cite{Kenter:2015vt}~learn semantic features for every sentence in the dataset based on a saliency weighted network for which the BM25 algorithm is used.
However, the features are being learned for every sentence prior to test time, and therefore not applicable in a real-time streaming context.
Finally, \cite{Kang:2014un}~calculate a cosine similarity matrix between the words of two sentences that are sorted based on their idf value, which they use as a feature vector for an MLP.
Their approach is similar to our work in the sense that the authors use idf information to rescale term contribution.
Their primary goal, however, is calculating semantic similarity instead of learning a sentence representation.
In fact, the authors totally discard the original word embeddings and only use the calculated cosine similarity features.

\section{Methodology}
\label{sec:methodology}
The core principle of our methodology is to assign a weight to each word in a short text.
These weights are determined based on the idf value of the individual words in that text.
The idea is that important words -- i.e.~words that are needed to determine most of the text's semantics -- usually have higher idf values than less important words, such as articles and auxiliaries\dots~Indeed, the latter appear more frequently in various different texts, while words with a high idf value mostly occur in similar contexts.
The final goal is to combine the weighted words into a semantically effective, single text representation.

To achieve this goal, we will model the problem of finding a suitable text representation as a semantic similarity task between couples of short texts.
In order to classify such couples of text fragments into either semantically related pairs or non-related pairs, the vector representations of these two texts are directly compared.
In this paper we use a simple threshold function on the distance between the two text representations, as we want related pairs to lie close to each other in their representation space, and non-related pairs to lie far apart:
\begin{align}
g(t_1, t_2) = \begin{cases} \text{pair} & \text{if } d(t_1, t_2) \leq \theta \\ \text{non-pair} & \text{if } d(t_1, t_2) > \theta \end{cases}.
\label{eq:decisionstump}
\end{align}
In this expression $t_1$ and $t_2$ are two short text vector representations of dimensionality $\nu$, $d\colon (x,y)\in \mathbb{R}^{2\nu} \rightarrow \mathbb{R}^+$ is a vector distance function of choice (e.g.~cosine distance, euclidean distance\dots), $\theta$ is a threshold, and $g(\cdot)$ is the binary prediction of semantic relatedness.

\subsection{Basic architecture}
\label{sec:architecture}
\begin{figure}[t!]
\centering
\begin{tikzpicture}[scale=0.97,inner sep=0pt]
	\node [draw=none, anchor=south west] () at (0,0) {\includegraphics[width=.9080\linewidth]{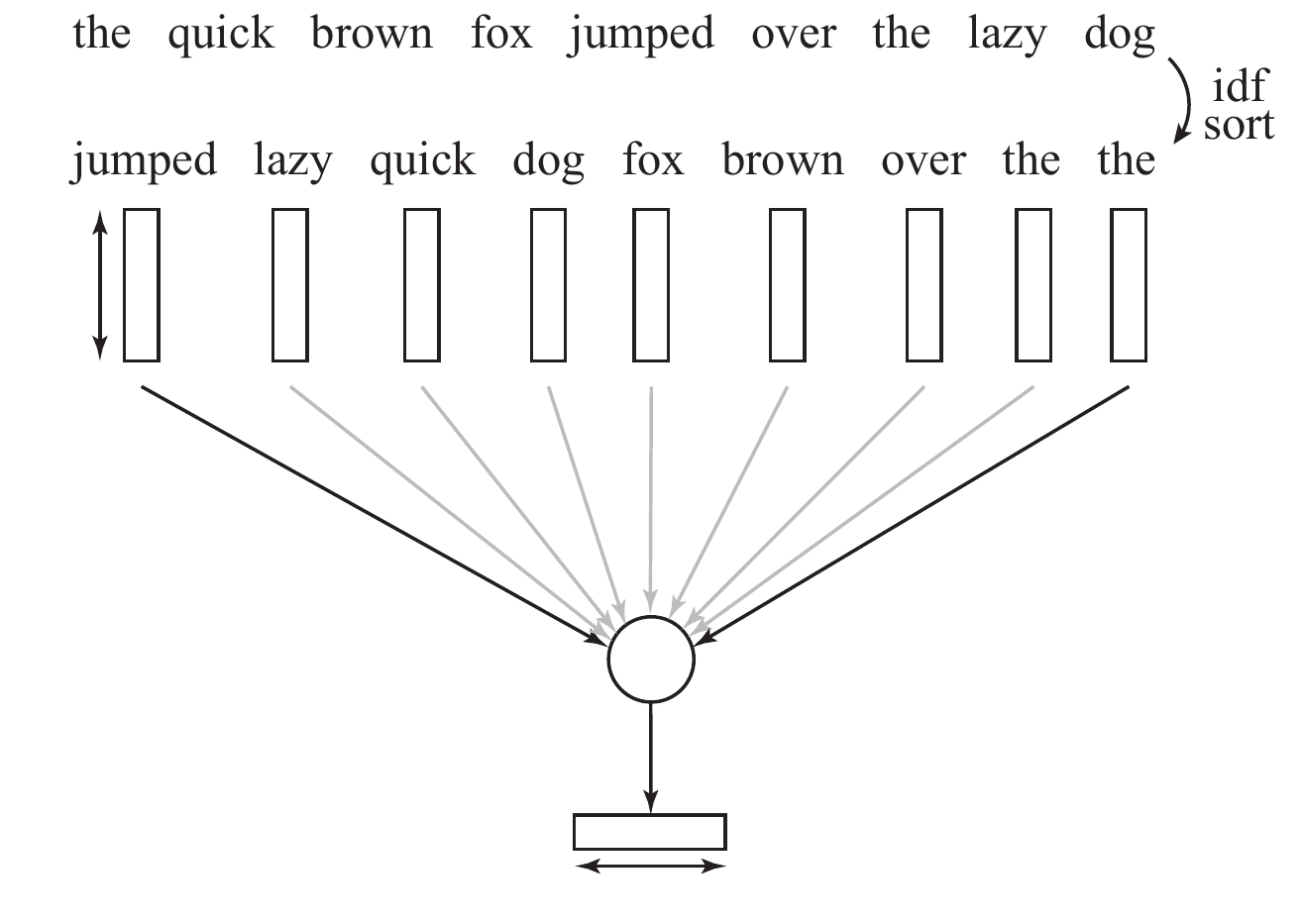}};
	
	\node[below=2pt] (bO) at (2.0,2.75) {\small $w_1$};
	\node[below=2pt] (b8) at (6.2,2.75) {\small $w_9$};
	
	\node[below=2pt] (d) at (4.19,2.75) {$\cdots$};
	 
	\node[below=2pt] (s) at (4.16,1.88) {$\Sigma$};
	
	\node[below=2pt] (n) at (4.36,1.30) {\small $\nicefrac{1}{9}$};
	
	\node[left=2pt] (bO) at (0.6,4.0) {\small $\nu$};
	\node[below=2pt] (bO) at (4.15,0.35) {\small $\nu$};
	
\end{tikzpicture}
\caption{Illustration of the weighted average approach for a toy sentence of nine words long and word vectors of dimension $\nu$.}
\label{fig:example_1}
\end{figure}
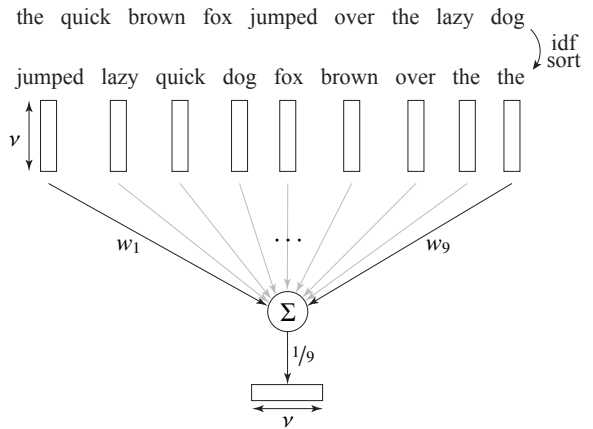

As mentioned before, we will assign a weight to each word in a text according to that word's idf value.
To learn these weights, we devise a model that is visualised in Figure \ref{fig:example_1}.
In the learning scheme, we use related and non-related couples of text as input data.
First, the words in every text are sorted from high to low idf values.
Original word order is therefore discarded, as is the case in usual standard aggregation operations.
After that, every embedding vector for each of the sorted words is multiplied with a weight that can be learned.
Finally, the weighted vectors are averaged to arrive at a single text representation.

In more detail, consider a dataset $\dataset$ consisting of couples of short texts.
An arbitrary couple is denoted by $C$, and the two texts of $C$ by $\texta$ and $\textb$.
We indicate the vector representation of word $j$ in text $C^\alpha$ by $\texta_j$.
All word vectors have the same dimensionality $\nu$.
Each text $\texta$ also has an associated length $n(\texta)$, i.e.~the number of words in $\texta$.
For now, in this section, we assume that $n \triangleq n(\texta) = n(\textb), \forall C \in \dataset$, a notion we will relax in Section \ref{sec:variablelength}.
The final goal of our methodology is to arrive at a vector representation for both texts in $C$, denoted by $t(\texta)$ and $t(\textb)$.
Denoting the sorted texts as $\sorteda$ and $\sortedb$, we arrive at a vector representation $t(\texta)$ and $t(\textb)$ through the following equation:
\begin{align}
\forall \ell \in \set{\alpha, \beta}\colon t(C^\ell) = \frac{1}{n} \sum_{j=1}^{n} w_j \cdot C^{\ell'}_j,
\label{eq:vectorrepresentation}
\end{align}
in which $w_j, j\in\{1,\dots,n\}$ are the weights to be learned.
As such, we create a weighted sum of the individual embeddings, the weights of which are only related to the rank order according to the idf values of all words in the fragment.

The model we construct through this procedure is related to the siamese neural network with parameter sharing from the early nineties (\cite{Bromley:1993wo}).
The learning procedure of such models is as follows.
We first calculate the vector representations for both texts in a particular couple through Equation \eqref{eq:vectorrepresentation}, both using the same weights, after which we compare the two vector representations through a loss function $\loss{t(\texta), t(\textb)}$ that we wish to minimize.
After that, the weights are updated through a gradient descent procedure using a learning rate $\eta$:
\begin{align}
\forall j \in \set{1,\dots,n}\colon w_j \leftarrow w_j - \eta \cdot \frac{\partial}{\partial w_j} \loss{t(\texta), t(\textb)}.
\end{align}

\subsection{Loss functions}
\label{sec:lossfunctions}

\begin{figure}[t!]
\centering
\begin{tikzpicture}[scale=.97,inner sep=0pt]
	\node [draw=none, anchor=south west] () at (0,0) {\includegraphics[trim = 2.5cm 1.5cm 1.5cm 1.5cm, clip, width=.79\linewidth]{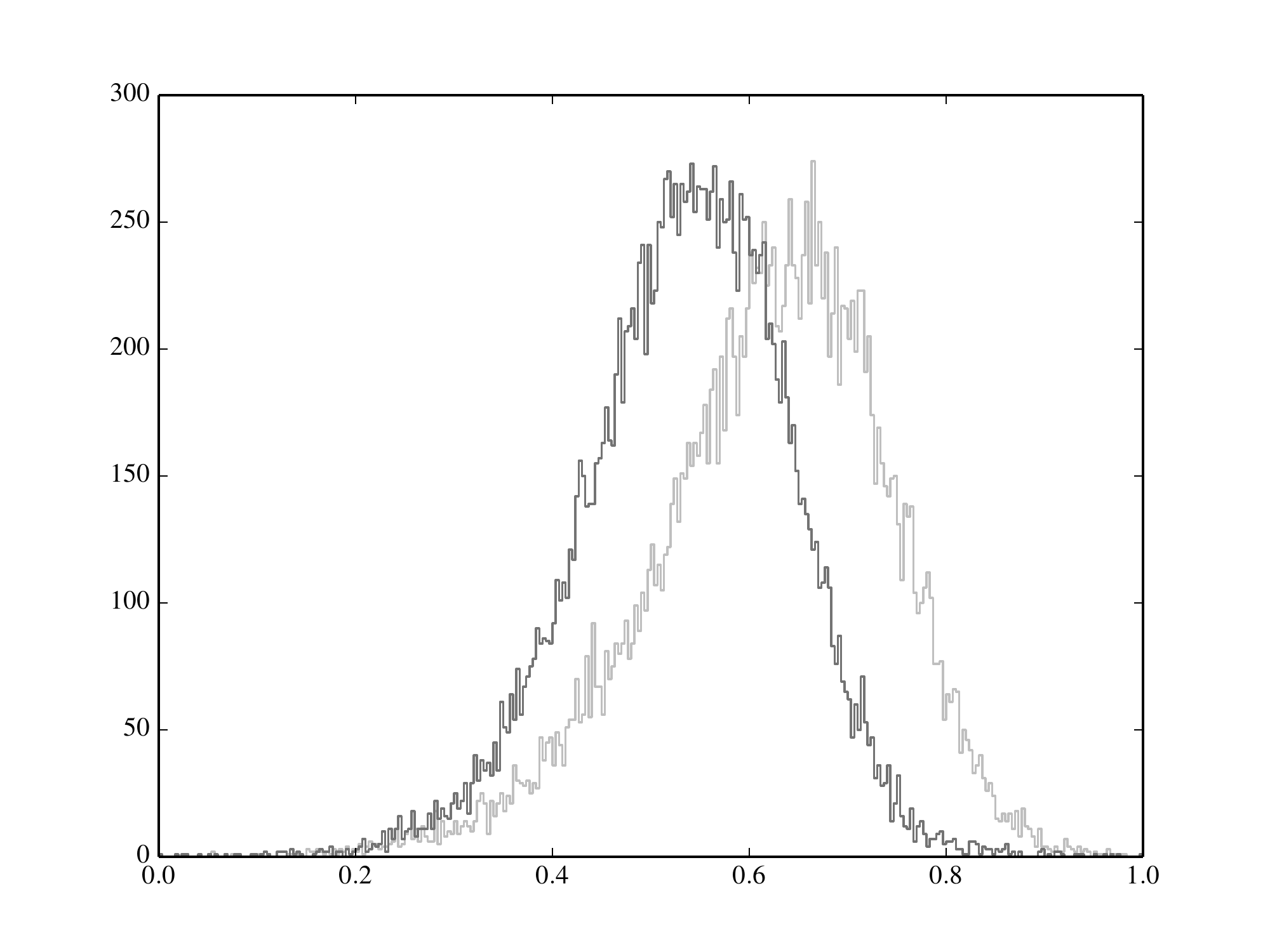}};
	\node[below=2pt] (bO) at (0.1,0.0) {$0$};
	\node[below=2pt] (b3) at (6.95,0.0) {$1$};
	\draw[draw opacity=0] (0.1,0.0) -- node[below=3pt] {Normalized distance} (6.95,0.0);
	\draw[draw opacity=0] (0.0, 0.1) -- node[left=3pt] {\rotatebox{90}{Number of couples}} (0.0, 5.3);
	\node[left=2pt] (lO) at (0.0,0.1) {$0$};
	\node[left=2pt] (l1) at (0.0,5.3) {$300$};
\end{tikzpicture}
\caption{Example distributions of distances between non-related pairs (light grey) and related pairs (dark grey) on Twitter data, using a simple mean of word embeddings.}
\label{fig:mean_distribution}
\end{figure}
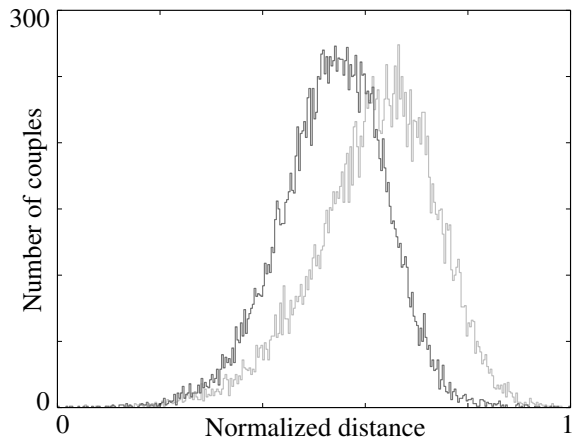

As pointed out in the beginning of this section, we want to have semantically related texts to lie close to each other in the representation space, and non-related texts to lie far apart from each other.
We can visually inspect the distribution of the distances between every couple in the dataset.
In fact, we calculate two distributions, one for the related pairs and one for the non-related pairs.
Two examples of such distributions, created using Twitter data and an average word embedding text representation, are shown in Figure \ref{fig:mean_distribution}.
Related pairs tend to lie closer to each other than non-related pairs.
There is however a considerable overlap between the two distributions. This means that making binary decisions on similarity based on a well-chosen distance threshold will lead to a large error.
The goal is to reduce this error, and thus to minimize the overlap between the two distributions.
Directly minimizing the overlap is difficult, since it requires the overlap as a function of the model weights.
We will instead describe two different loss functions as an approximation to this problem.

The first loss function is related to the contrastive loss function regularly used in siamese neural architectures (\cite{Hadsell:2006}). We define the quantity $p_C$ as follows:
\begin{align}
p_C \triangleq \begin{cases} 1 & \text{if } C\text{ is a related pair,} \\ -1 & \text{if } C\text{ is a non-related pair.} \end{cases}
\end{align}
The loss function, which we will conveniently call the contrastive-based loss, is then given by:
\begin{align}
\closs{t(\texta), t(\textb)} = p_C \cdot d\left(t(\texta), t(\textb)\right),
\end{align}
in which $d\colon (x,y)\in \mathbb{R}^{2\nu} \rightarrow \mathbb{R}^+$ is a vector distance function of choice, as before.
Note that, when trying to minimize this loss, related pairs will get pushed to each other in the representation space, while non-related pairs will get dragged apart.

This loss function, however, has two main problems.
First, there is an imbalance between the loss for related pairs and non-related pairs, in which the latter can get an arbitrarily negative loss, while the related pairs' loss cannot be pushed below zero.
To solve this, we could add a maximum possible distance -- which is e.g.~1 in the case of cosine distance -- but this cannot be generalized to arbitrary distance functions.
Second, this loss function can skew distance distributions, so that minimizing overlap between distributions is not guaranteed.
In fact, the overlap can even increase while minimizing this loss.
This happens, for example, when the distance between some of the related pairs can be drastically reduced while other related pairs get dragged farther apart, and vice versa for the non-related pairs.
The loss function as such allows this to happen, since it can focus on data points that are easier to shift towards or away from each other and it can ignore what happens to the other data points.
As long as the contribution of these shifts to the overall loss remains dominant, the loss will diminish, although the predictions according to Equation \eqref{eq:decisionstump} will be worse.
Despite these problems we still consider this loss function due to its simplicity.
The derivative with respect to weight $w_j$ is given by:
\begin{align}
\frac{\partial}{\partial w_j} &\closs{t(\texta), t(\textb)} \nonumber\\
&= \frac{p_C}{n} \left.\Big(\nabla d(x,y)\Big)\right\vert_{x=\frac{1}{n} \sum_{j=1}^{n} w_j \sorteda_j, y=\frac{1}{n} \sum_{j=1}^{n} w_j \sortedb_j} \begin{bmatrix}
\sorteda_j\\ 
\sortedb_j
\end{bmatrix}.
\label{eq:contrastivegradient}
\end{align}

In a second loss function we try to mitigate the loss imbalance and potential skewing of the distributions caused by the contrastive-based loss function.
For this purpose we will use the median, as it is a very robust statistic insensitive to outliers.
As we need multiple data points to calculate the median, this loss function can only be used in the context of minibatch gradient descent, in which the number of positive and negative examples in each minibatch is balanced.
In practice we consider a minibatch $B\subset\dataset$ of $n(B)$ randomly sampled data points, in which there are exactly $\nicefrac{n(B)}{2}$ related pairs and $\nicefrac{n(B)}{2}$ non-related pairs.
We consider the couple of texts $M(B) \in B$ as the median couple if $\mu(B) = d(M(B)^\alpha, M(B)^\beta)$ is the median of all distances between the couples in $B$:
\begin{align}
M(B)   &\triangleq \arg\median_{C\in B} d(C^\alpha, C^\beta);\\
\mu(B) &\triangleq \median_{C\in B} d(C^\alpha, C^\beta).
\end{align}
Since minibatch $B$ is randomly sampled, we can consider $\mu(B)$ as an approximation to the optimal split point between related and non-related pairs, in the sense of threshold $\theta$ in Equation \eqref{eq:decisionstump}.
We thus consider all related pairs with a distance larger than $\mu(B)$, and all non-related pairs with a distance smaller than $\mu(B)$ to be classified incorrectly.
Since minimizing a 0-1 loss is NP-hard, we use a scaled cross-entropy function in our loss, which we will call the median-based loss:
\begin{align}
&\mloss{t(\texta), t(\textb), B} \nonumber\\
&\qquad = \ln\left[1 + \exp\bigg(-\kappa p_C\Big(\mu(B) - d\big(t(\texta), t(\textb)\big)\Big)\bigg)\right],
\end{align}
in which $\kappa$ is a hyperparameter. The derivative with respect to weight $w_j$ is given by the following expression (in which $\sigma(\cdot)$ is the sigmoid function):
\begin{align}
&\frac{\partial}{\partial w_j} \mloss{t(\texta), t(\textb), B} \nonumber\\
&= \kappa\sigma\left(-\kappa p_C\left(\mu(B) - d\left(t(\texta), t(\textb)\right)\right)\right) \nonumber\\
&\quad \cdot\frac{\partial}{\partial w_j} \left(\closs{t(\texta), t(\textb)} - \closs{t(M(B)^\alpha), t(M(B)^\beta)}\right).
\end{align}

\subsection{Texts with variable length}
\label{sec:variablelength}

The method described thus far is only applicable to short texts of a fixed length, which is limiting.
In this section we will extend our technique to texts of a variable, but given maximum length.
For this purpose we have devised multiple approaches, of which we will elaborate the one that performed best in the experiments.

Suppose that all texts have a fixed maximum length of $n_{\max}$.
In the learning procedure we will learn a total of $n_{\max}$ weights with the techniques described earlier.
To find the weights for a text with length $m \leq n_{\max}$ we will use subsampling and linear interpolation.
That is, for an arbitrary text $C^{\ell}$ we first find the sequence of real-valued indices $I(C^{\ell}, j), j\in\set{1,\dots,n(C^{\ell})}$ through subsampling:
\begin{align}
\forall j\in\set{1,\dots,n(C^{\ell})}\colon I(C^{\ell}, j) \triangleq 1 + \frac{(j-1) (n_{\max} - 1)}{n(C^{\ell}) - 1}.
\end{align}
Then, in the second step, we calculate the new weights $z_j(C^{\ell}), j\in\set{1,\dots,n(C^{\ell})}$ for the words in $C^{\ell}$ through linear interpolation, in which $\epsilon$ is arbitrarily small:
\begin{align}
&\forall j \in \set{1,\dots, n(C^{\ell})}\colon z_j(C^{\ell}) \triangleq \nonumber\\
&\quad\quad\frac{(w_{\ceil{I_j(C^{\ell})}} - w_{\floor{I_j(C^{\ell})}})(I(C^{\ell}, j) - \floor{I_j(C^{\ell})})}{\ceil{I_j(C^{\ell})} - \floor{I_j(C^{\ell})} + \epsilon} + w_{\ceil{I_j(C^{\ell})}}.
\end{align}
In this, $\ceil{\cdot}$ and $\floor{\cdot}$ are resp.~the ceil and floor functions.
Finally, Equation \eqref{eq:vectorrepresentation} needs to be updated with these new weights instead of $w_j$:
\begin{align}
\forall \ell \in \set{\alpha, \beta}\colon t(C^\ell) = \frac{1}{n(C^{\ell'})} \sum_{j=1}^{n(C^{\ell'})} z_j(C^{\ell'}) \cdot C^{\ell'}_j.
\end{align}
Calculating the derivative of the new weights with respect to the original weights is straightforward.

\section{Data collection}
\label{sec:datacollection}
To train the weights for the individual embeddings and to conduct experiments, we collect data from different sources.
First we gather textual pairs from Wikipedia which we will use as a base dataset to finetune our methodology.
We will also use this dataset to test our hypotheses and perform initial experiments.
The second dataset will consist of Twitter message pairs, which we will use to show that our method can be used in a practical streaming context.

\subsection{Wikipedia}
\label{sec:wikipedia}
We will perform our initial experiments in a base setting using English Wikipedia articles.
The most important benefit of using Wikipedia is that there is a lot of well structured data.
It is therefore a good starting point to collect a ground truth to finetune our methodology.

We use the English Wikipedia dump of March 4th 2015, and we remove its markup and punctuation.
We convert all letters to lower case and every number is replaced by a single character `0' (zero).
Next we construct related pairs of texts which both have the same, fixed length $n$.
To do this, we take a Wikipedia article and we extract $n$ consecutive words out of a paragraph.
Then we skip two words, after which we extract the next $n$ consecutive words, as long as they remain in the same paragraph.
To extract non-related text pairs we follow the same procedure, but we make sure that the two texts are from different articles, which we choose at random.
This approach is closely related to the data collection practice used in (\cite{Hu:2014uo}).
We want to emphasize again that our vision of semantic similarity is one of topic-based similarity instead of paraphrase-similarity, as discussed in the introduction.
This notion is reflected in our data collection.
We extract a total of 4.9 million related pairs and 4.9 million non-related pairs, each for fixed-length texts of 20 words long.
We also extract 4.9 million related and non-related pairs of which the texts varies in length between 10 and 30 words.
All datasets are divided into a train set of 1.5 million pairs, a test set of 1.5 million pairs and a validation set of 1.9 million pairs.

\subsection{Twitter}
\label{sec:twitter}
Twitter is a very different kind of medium than Wikipedia.
Wikipedia articles are written in a formal register and mostly consist of linguistically correct sentences.
Twitter messages, on the other hand, count no more than 140 characters and are full of spelling errors, abbreviations and slang.

We propose that two tweets are semantically related if they are generated by the same event.
As in (\cite{DeBoom:2015ur}), we require that such an event is represented by one or more hashtags.
Since Twitter posts and associated events are very noisy by nature, we restrict ourselves to tweets by 100 English news agencies.
We manually created this list through inspection of their Twitter accounts; the list is available through our GitHub page, see Section \ref{sec:conclusion}.

We gathered 341 949 tweets from all news agencies through the Twitter REST API at the end of August 2015.
We convert all words to lowercase, replace all numbers by the single character `0' and remove non-informative hashtags such as \textit{\#breaking}, \textit{\#update} and \textit{\#news}.
To generate related pairs out of these tweets, we consider four simple heuristic rules:
\begin{enumerate}
\item The number of words in each tweet, different from hashtags, mentions or URLs, should be at least 5.
\item The Jaccard similarity between the set of hashtags in both tweets should be at least 0.5.
\item The tweets should be sent no more than 15 minutes from each other.
\item The Jaccard similarity between the set of words in both tweets should be less than 0.5.
\end{enumerate}
We add the last rule in order to have sufficient word dissimilarity between the pairs, as tweets that mostly contain the same words are too easy to relate.
To generate non-related pairs, we remove rule 3 and rule 2 is changed: the Jaccard similarity between sets of hashtags should now be zero.
Using these heuristics, we generate a train set of 15 000 pairs, a validation set of 20 000 pairs and a test set of 13 645 pairs, of which we remove all overlapping hashtags.
We manually label 200 generated pairs and non-pairs, and we achieve an error rate of 28\%.
Due to the used heuristics and the linguistic nature of tweets in general, the ground truth can be considered very noisy; achieving an error rate lower than around 28\% on this dataset will therefore be difficult, and the gain would not lead to a better model of the human notion of similarity anyway.

\section{Experiments}
\label{sec:experiments}
In this section we discuss the results of several experiments on all aspects of our methodology given the data we collected.
First we will discuss some results using the Wikipedia dataset, after which we also take a look at the Twitter dataset.
We will use two performance metrics in our evaluation.
The first is the optimal split error, i.e.~we classify all pairs according to Equation \eqref{eq:decisionstump} -- after determining the optimal split point $\theta$ -- and we determine the classification error rate.
A second performance metric is the Jensen-Shannon (JS) divergence.
This is a symmetric distance measure between two probability distributions, related to the -- well-known, but asymmetric -- KL divergence.
We will use it to capture the difference between the related and non-related pairs' distributions, as shown in Figure \ref{fig:mean_distribution}.
The bigger the JS divergence, the greater the difference between the two distributions.

In our experiments we will use Google's \textit{word2vec} software to calculate word embeddings.
We choose Skip-gram with negative sampling as the learning algorithm, using a context window of five words and 400 dimensions.
We feed an entire cleaned English Wikipedia dump of March 4th 2015 to the algorithm, after which we arrive at a total vocabulary size of 2.2 million words.
Since we also need document frequencies, we calculate these for each of the vocabulary words using the same Wikipedia dump.

In previous work we showed that using an Euclidean distance function leads to a much better separation between related and non-related pairs than the more often used cosine distance, so we will also use the Euclidean distance throughout our experiments here (\cite{DeBoom:2015gl}).
Calculating the gradient of this distance function -- which is used in Equation \eqref{eq:contrastivegradient} -- is straightforward.

To obtain the results hereafter, we use the following procedure.
We use the train set to train the weights $w_j$ in Equation \eqref{eq:vectorrepresentation}.
The validation set is used to determine the optimal split point $\theta$ in Equation \eqref{eq:decisionstump}.
Finally predictions and evaluations are made on the test set.
In the next two subsections we discuss the results on the Wikipedia and Twitter datasets.


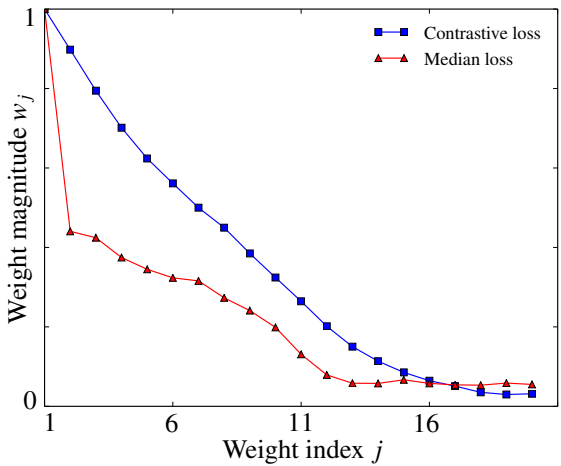
\begin{figure}[t!]
\centering
\begin{tikzpicture}[scale=.97,inner sep=0pt]
	\node [draw=none, anchor=south west] () at (0,0) {\includegraphics[trim = 2.5cm 1.5cm 1.5cm 1.5cm, clip, width=.79\linewidth]{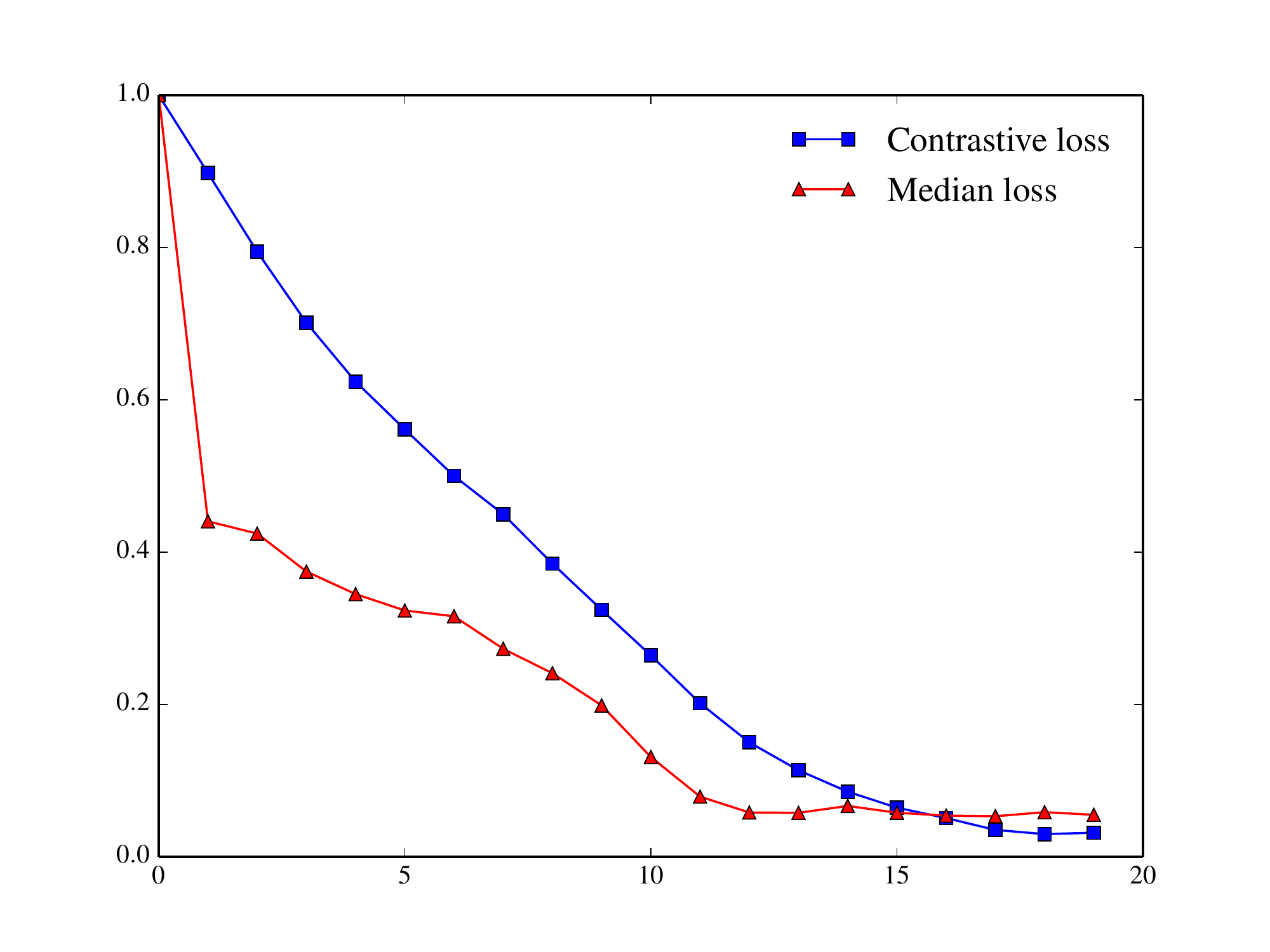}};

	\node[below=2pt] (bO) at (0.1,0.0) {$1$};
	\node[below=2pt] (b1) at (1.75,0.0) {$6$};
	\node[below=2pt] (b2) at (3.5,0.0) {${11}$};
	\node[below=2pt] (b3) at (5.2,0.0) {${16}$};
	\draw[draw opacity=0] (0.1,0.0) -- node[below=12pt] {Weight index $j$} (6.95,0.0);
	\draw[draw opacity=0] (0.0, 0.1) -- node[left=3pt] {\rotatebox{90}{Weight magnitude $w_j$}} (0.0, 5.3);
	\node[left=2pt] (lO) at (0.0,0.1) {$0$};
	\node[left=2pt] (l1) at (0.0,5.3) {$1$};
\end{tikzpicture}
\caption{Plot of the learned weight magnitudes for texts of 20 words long.}
\label{fig:weight_plot}
\end{figure}

\begin{table*}[t!]
\small
\centering
\caption{Results for the Wikipedia data, for texts of length 20 and variable lengths, using Wikipedia and Google embeddings.}
\label{table:results-wiki}
\begin{tabular}{l | c c | c c | c c}
\toprule
					& \multicolumn{2}{c|}{Wikipedia, 20 words} & \multicolumn{2}{c|}{Wikipedia, variable length} & \multicolumn{2}{c}{Google, variable length}\\
					& Split error & JS divergence & Split error & JS divergence & Split error & JS divergence\\
\hline
Tf-idf				& 19.60\% & 0.3698 & 22.23\% & 0.3190 & \textbf{22.50\%} & \textbf{0.3130}\\
\hline
Mean					& 19.43\% & 0.3781 & 25.61\% & 0.2543 & 34.41\% & 0.1130\\
Max					& 19.05\% & 0.3981 & 27.09\% & 0.2308 & 37.05\% & 0.0842\\
Min/max 				& 16.78\% & 0.4520 & 26.19\% & 0.2487 & 35.82\% & 0.1032\\
\hline
Mean, top 30\% idf		& 17.02\% & 0.4435 & 22.67\% & 0.3182 & 32.07\% & 0.1512\\
Max, top 30\% idf		& 18.05\% & 0.4193 & 28.32\% & 0.2150 & 36.38\% & 0.0999\\
Min/max, top 30\% idf	& 16.40\% & 0.4581 & 27.86\% & 0.2260 & 35.61\% & 0.1140\\
Mean, idf-weighted 		& 24.00\% & 0.2767 & 27.51\% & 0.2139 & 33.88\% & 0.1185\\
\hline
Learned weights, contrastive 	& 14.44\% & 0.5080 & 21.20\% & 0.3503 & 30.50\% & 0.1776\\
Learned weights, median 	& \textbf{14.06\%} & \textbf{0.5184} & \textbf{16.41\%} & \textbf{0.4602} & 25.10\% & 0.2641\\
\bottomrule
\end{tabular}
\end{table*}

\subsection{Baselines}
We will compare the performance of our methods to several baseline mechanisms that construct sentence vector representations.
The simplest and most widely used baseline is a tf-idf vector.
Comparing two tf-idf vectors is done through a standard cosine similarity.
In a second baseline we simply take, for every dimension, the mean across all embeddings:
\begin{align}
\label{eq:baseline-max}
\forall \ell \in \set{\alpha, \beta}\colon \forall k\in\set{1,\dots,\nu}\colon t(C^\ell)_k = \mean_{j} C^{\ell}_{j, k}.
\end{align}
In the third baseline we replace the mean by a maximum operation.
Taking a mean or a maximum is a very common approach to combine word embeddings -- or other intermediary representations in an NLP architecture -- into a sentence representation.
We can also replace the maximum in Equation \eqref{eq:baseline-max} by a minimum.
The fourth baseline is a concatenation of the maximum and minimum vectors, resulting in a vector having two times the original dimensionality (`min/max').
We can also apply the three previous baselines -- i.e. `mean', `max' and `min/max' -- only considering words with a high idf value (`top 30\% idf').
That is, we sort the words in a text based on their idf values, and we take the mean or maximum of the top 30\%.
In a final baseline we weigh each word in the sentence with its corresponding idf value and then take the mean (`mean, idf-weighted').

\subsection{Details on the learning procedure}
In the results below we use the methodology from Section \ref{sec:methodology} to learn weights for individual words in a short text.
All procedures are implemented with the Theano\footnote{\url{deeplearning.net/software/theano}} framework and executed using an Nvidia Tesla K40 GPU.
We use mini-batch stochastic gradient descent as learning procedure and $L_2$ regularization on the weights.
The total loss for one training batch thus becomes:
\begin{align}
\mathcal{L}_{\text{total}} = \mathcal{L}_{\text{batch}} + \lambda \sum_{j=1}^{n_{\max}} w_j^2.
\end{align}
In this, we empirically set parameter $\lambda$ to 0.001, and $\mathcal{L}_{\text{batch}}$ is either equal to the contrastive or median-based loss depending on the experiment.
The batch size is equal to 100 text couples, of which 50 are related pairs and 50 are non-related pairs.
An initial learning rate $\eta$ of 0.01 is employed, which we immediately lower to 0.001 once the average epoch loss starts to deteriorate.
After that, we stop training when the loss difference between epochs is below 0.05\%.
The weights are initialized uniformly to a value of 0.5.
The entire procedure is visualised in Algorithm \ref{algo:training}.
To determine the optimal value of the hyperparameter $\kappa$ in the median-based loss, we use five-fold cross-validation and a grid search procedure.

\begin{algorithm}
\DontPrintSemicolon
  $\forall j\colon w_j \leftarrow 0.5$\;
  $\eta \leftarrow 0.01$\;
  $\mathcal{L}_{\text{mean-old}} \leftarrow \infty$\;
  \Repeat{STOP}{
  $\mathcal{L}_{\text{mean}} = 0$\hfill\CommentSty{~~~~~~~~~~~~~~~~~~~~~~~$\triangleright$new epoch}\;
  \For{batch $i \in$ dataset}{
  		$\mathcal{L}_{\text{total}} \leftarrow \mathcal{L}_{\text{batch}_i} + \lambda \sum_{j=1}^{n_{\max}} w_j^2$\;
  		$\forall j\colon w_j \leftarrow w_j - \eta \cdot \frac{\partial}{\partial w_j} \mathcal{L}_{\text{total}}$\hfill\CommentSty{$\triangleright$gradient descent}\;
  		$\mathcal{L}_{\text{mean}} \leftarrow \frac{(i-1)\mathcal{L}_{\text{mean}} + \mathcal{L}_{\text{total}}}{i}$\hfill\CommentSty{~~~~$\triangleright$update mean loss}\;
  	}
  	\If{$\eta > 0.001 \land \mathcal{L}_{\text{mean}} > \mathcal{L}_{\text{mean-old}}$}{
  		$\eta \leftarrow 0.001$\;
  	}
  	\If{$\eta == 0.001 \land \mathcal{L}_{\text{mean-old}} - \mathcal{L}_{\text{mean}} < 0.0005$}{
  		STOP\;
  	}
  	$\mathcal{L}_{\text{mean-old}} \leftarrow \mathcal{L}_{\text{mean}}$\;
  }
  \caption{Detailed training procedure}\label{algo:training}
\end{algorithm}

\subsection{Results on Wikipedia}


We train weights for Wikipedia texts with a fixed length of 20 words using the training procedure described in the previous section.
The weights already converge before the first training epoch is finished, that is, without having seen all examples in the train set.
This is due to the simplicity of our model -- i.e.~there are only limited degrees of freedom to be learned -- and the large train set.
Through cross-validation and grid-search we find an optimal value for $\kappa = 160$ used in the median-based loss.
The resulting weights for texts of 20 words long are visualized in Figure \ref{fig:weight_plot}, for both the contrastive- and median-based loss.
In both cases, the weights drop in magnitude with increasing index; this confirms the hypothesis that words with a low idf contribute less to the overall meaning of a sentence than high idf words.
For the median-based loss, the first word is clearly the most important one, as the second weight is only half the first weight.
It is important to point out that we observe a similar monotonically decreasing pattern for texts of 10 words and 30 words long, which means that we use an equal proportion of important to less important words, no matter how long the sentence is.
From the 16th word on, the weights are close to zero, so these words almost never contribute to the meaning of a text.
In comparison, there are, by experiment, eight non-informative words on average in a text of twenty words long.

The results of the experiments are summarized in Table \ref{table:results-wiki}.
In a first experiment we compare the performance of our approach to all baselines for texts of length 20 and with word embeddings trained on Wikipedia.
Our method significantly outperforms all baselines ($p < 0.001$, two-tailed binomial test) by a high margin for both losses.
We also see that a plain tf-idf vector can compete with the simplest and most widely used baselines.
We further notice that concatenating the minimum and maximum vectors performs approx 2.25\% better than when using a maximum vector alone, which implies that the sign in word embeddings holds semantically relevant information.

In a second experiment we vary the length of the texts between 10 and 30 words.
The overall performance drop can be addressed to the presence of text pairs with a length shorter than 20 words.
Tf-idf now outperforms all baselines.
For texts of 30 words the probability of word overlap is after all much higher than for texts of 10 words long; pairs of long texts thus help lower the error rate for tf-idf.
Our method is still the overall best performer ($p < 0.001$, two-tailed binomial test), but this time the median-based loss is able to improve the contrastive-based loss by a high margin of almost 5\% in split error, showing that the former loss is much more robust in a variable-length setting.

Finally, we also performed experiments with word embeddings from Google News, see footnote \ref{footnote:word2vec}.
We want to stress that we did not retrain the weights for this experiment in order to test whether our technique is indeed applicable out-of-the-box.
There is only 20.6\% vocabulary overlap between the Wikipedia and Google word2vec model, and there are only 300 dimensions instead of 400, which together can explain the overall performance drop of the word embedding techniques.
It is also possible that the Google embeddings are no right fit to be used on Wikipedia texts.
Although our model was not trained to perform well on Google embeddings, we still achieve the best error rate of all embedding baselines ($p < 0.001$, two-tailed binomial test), and again the median loss outperforms the contrastive loss by approx 5\%.
Tf-idf, on the other hand, is the clear winner here; it did almost not suffer from the vocabulary reduction.
It shows that vector dimensionality and context of usage are important factors in choosing or training word embeddings.

\subsection{Results on Twitter}

Next we perform experiments on the data collected from Twitter.
We do not train additional word embeddings for Twitter, but we keep using the Wikipedia embeddings, since we have restricted ourselves to tweets of news publishers, who mainly use clean language.
We also keep the same setting for $\kappa$ as in the Wikipedia experiments.

The results for the Twitter experiments are shown in Table \ref{table:results-tweets}.
As expected, the error rate is quite high given the noise present in the dataset.
We also notice that the split error remains slightly higher than the human error rate of 28\%.
Tf-idf performs worst in this experiment.
Compared to Wikipedia, tf-idf vectors for tweets are much sparser, which leads to a higher error rate.
Tf-idf is clearly not fit to represent tweets efficiently.
The baselines on the other hand have a much better, but overall comparable performance.
Our method with median-based loss performs the best.
The approach using contrastive loss performs worst among all embedding baselines, as during training the distribution of distances between related and between non-related texts rapidly gets skewed and develops additional modes.
This causes the overall training loss to decrease, while the overlap between the related pairs' and non-related pairs' distribution further increases.
The overall improvement of the median-based loss over the idf-weighted baseline is not statistically significant ($p > 0.05$, two-tailed binomial test); so, based on this Twitter dataset alone, we cannot draw any statistically sound conclusion whether our method is better in terms of split error than the idf-weighted baseline.
Combined with the results from Table \ref{table:results-wiki}, however, we can conclude that choosing median-based learned weights is generally recommended.
\begin{table}[t!]
\small
\centering
\caption{Results for the Twitter data using Wikipedia embeddings.}
\label{table:results-tweets}
\begin{tabular}{l | c c}
\toprule
					& Split error & JS divergence \\
\hline
Tf-idf				& 43.09\% & 0.0634 \\
\hline
Mean					& 33.68\% & 0.0783 \\
Max					& 34.85\% & 0.0668 \\
Min/max 				& 33.80\% & 0.0734 \\
\hline
Mean, top 30\% idf		& 32.60\% & 0.0811 \\
Max, top 30\% idf		& 33.38\% & 0.0740 \\
Min/max, top 30\% idf	& 32.86\% & 0.0762 \\
Mean, idf-weighted 		& 31.28\% & 0.0886 \\
\hline
Learned weights, contrastive & 35.48\% & 0.0658 \\
Learned weights, median & \textbf{30.88\%} & \textbf{0.0900} \\
\bottomrule
\end{tabular}
\end{table}

\section{Conclusion}
\label{sec:conclusion}
We devised an effective method to derive vector representations for very short fragments of text.
For this purpose we learned to weigh word embeddings based on their idf value, using both a contrastive-based loss function and a novel median-based loss function that can effectively mitigate the effect of outliers.
Our method is applicable to texts of a fixed length, but can easily be extended to texts of a variable length through subsampling and linear interpolation of the learned weights.
Our method can be applied out-of-the-box, that is, there is no need to retrain the model when using different types of word embeddings.
We showed that our method outperforms widely-used baselines that naively combine word embeddings into a text representation, using both toy Wikipedia and real-word Twitter data. All code for this paper is readily available on our GitHub page \url{github.com/cedricdeboom/RepresentationLearning}.

\section*{Acknowledgments}
This work is soon to be published in Pattern Recognition Letters.
Cedric De Boom is funded by a Ph.D.~grant of the Flanders Research Foundation (FWO). Steven Van Canneyt is funded by a Ph.D.~grant of the Agency for Innovation by Science and Technology in Flanders (IWT). We acknowledge Nvidia for its generous hardware support.

\balance

\bibliographystyle{model2-names}

\end{document}